\documentclass[12pt,preprint]{aastex}
\usepackage{natbib}
\slugcomment{submitted to AJ, \today}

\begin{document}

\title{The size, density, and formation of the Orcus-Vanth system in the Kuiper belt}
\author{M.E. Brown\altaffilmark{1}, D. Ragozzine\altaffilmark{1,2}, J. Stansberry\altaffilmark{3}, W.C. Fraser\altaffilmark{1}
\altaffiltext{1}{Division of Geological and Planetary Sciences, California Institute
of Technology, Pasadena, CA 91125}
\altaffiltext{2}{now at Harvard-Smithsonian Center for Astrophysics, Cambridge, MA 02138}
\altaffiltext{3}{Steward Observatory, University of Arizona, Tucson, AZ 08544}}
\email{mbrown@caltech.edu}

\begin{abstract}
The Kuiper belt object Orcus and its satellite Vanth form an unusual system
in the Kuiper belt. While most large Kuiper belt objects have small satellites
in circular orbits \citep{2008ssbn.book..335B} and smaller Kuiper belt objects and their
satellites tend to be much closer in size \citep{2008ssbn.book..345N}, 
Orcus sits in
between. Orcus is amongst the largest objects known in the Kuiper belt, 
but the relative size of Vanth is much larger than that of the tiny satellites
of the other large objects \citep{2008ssbn.book..335B}. 
Here we characterize the physical
and orbital characteristics of the Orcus-Vanth system in an attempt
to distinguish discuss possible
formation scenarios. From {\it Hubble Space Telescope} observations
we find that Orcus and Vanth have different visible colors and that
Vanth does not share the water ice absorption feature seen in the
infrared spectrum of Orcus. We also find
that Vanth has a nearly face-on 
circular orbit with a period of $9.5393\pm0.0001$ days and
semimajor axis of $8980\pm20 $km, implying a system mass of $6.32\pm 0.01
\times 10^{20}$ kg or 3.8\% the mass of dwarf planet Eris. 
From {\it Spitzer Space Telescope} observations
we find that the thermal emission is consistent with a single body
with diameter 
$940\pm70$ km and a geometric albedo of $0.28\pm0.04$. Assuming equal
densities and albedos, this measurements implies sizes of Orcus
and Vanth of 900 and 280 km, respectively, and a mass ratio of 33.
Assuming a factor of 2 lower albedo for the non-icy Vanth, however,
implies sizes of 820 and 640 km and a mass ratio of 2. 
The measured density depends on the assumed albedo ratio of the
two objects but is approximately $1.5\pm0.3$ g cm$^{-3}$, midway between 
typical densities measured for larger and for smaller objects.
The orbit and mass
ratio is consistent with formation from a giant impact and subsequent
outward tidal evolution and even consistent with the system having
now achieved a double synchronous state. Because of the large angle
between the plane of the heliocentric orbit of Orcus and the plane of the 
orbit of Vanth, the system can equally well
be explained, however, by initial eccentric capture, Kozai cycling which
increases the eccentricity and decreases the pericenter of the orbit
of Vanth, and subsequent tidal evolution inward. We discuss implications
of these formation mechanisms.
\end{abstract}

\section{Introduction}
Orcus is among the brightest known bodies in Kuiper belt and appears 
unique in several ways. Its reflectance spectrum shows the deepest 
water ice absorption of any Kuiper belt object (KBO) that is not
associated with the Haumea collisional family 
\citep{2005A&A...437.1115D,2005ApJ...627.1057T,2008A&A...479L..13B}.
Compositionally, Orcus appears in 
many ways to be at a transitional size between the moderately
common medium-sized Kuiper belt objects (KBOs) which tend to be
spectrally bland \citep{2008AJ....135...55B,2009Icar..201..272G} 
and the rare large Kuiper belt objects 
which are massive enough to retain volatiles such as methane on their
surfaces \citep{2007ApJ...659L..61S}. In addition, Orcus has one of the 
brightest satellites among the large KBOs \citep{2008ssbn.book..335B}, relative 
to the primary. The large fraction
of small satellites around large KBOs 
led Brown et al. (2005) to suggest that many large
KBOs suffered collisions which led to satellite formation. In contrast, the
similar brightnesses and large angular momenta
of the binary members of other KBO systems has been used
to suggest a capture origin for these objects \citep{2002Natur.420..643G,2008ssbn.book..345N}. Again, 
Orcus appears to be near the boundary of these two regimes.

Another dichotomy exists between the larger and the smaller KBOs that have
satellites. While Eris, Pluto, and Haumea -- which all have small presumably
collisionally formed satellites --  have densities 
$\sim$2 g cm$^{-3}$ or higher \citep{2007Sci...316.1585B, 2006AJ....132..290B,2006ApJ...639.1238R}, smaller Kuiper belt binaries
have densities of 1 g cm$^{-3}$ and even
lower \citep{2006ApJ...643..556S, 2008Icar..197..260G}. Orcus, being
intermediate in size between the two populations, could help to shed insight
on this apparent bifurcation.

In order to investigate the properties of the Orcus system, we obtained Hubble 
Space Telescope (HST) observations to determine the orbit, and optical
and infrared colors of the components of the system, and Spitzer Space Telescope
observations of thermal emission in order to determine the
size and thus density of the system.

\section{The orbit of Vanth}
Vanth\footnote{note to editor/reviewer: Vanth is the name, as yet unofficial,
proposed to the IAU for the satellite},
the satellite of Orcus, was discovered in observations from the High 
Resolution Camera (HRC) of the Advanced Camera for Surveys (ACS) on HST 
obtained on 13 Nov 2005. 
Followup observations were obtained a year later in order
to determine the orbit of the system. In each HST orbit, 
8 exposures were obtained
in the F606W filter with 275 second exposures.

In order to determine accurate astrometric positions of the 
satellite relative to
Orcus, we performed detailed point spread function 
(PSF) fitting of the system.
A five-times oversampled theoretical PSF was constructed for 
the approximate pixel location of Orcus on the HRC 
image using the HST PSF modeling
software TinyTim, and a least-squares fit was 
performed optimizing the sub-pixel centers of Orcus and the satellite, the total flux of Orcus and the satellite, and the
magnitude of the sky background. The uncertainties for each observation are
determined from the scatter within a single HST orbit. In some cases it appears
that individual observations within an orbit detect consistent motion on minutes-long
time scales, nonetheless we assume that all deviation within one orbit is due to
measurement error.
Observations of the Orcus system were also obtained with the Wide Field/Planetary Camera 2 (WFPC2) on HST as well as NICMOS, the infrared camera. Astrometric positions
of the satellite were determined from these observations identically, using 
appropropriate PSFs for each camera and filter. All astrometric observations are
shown in Figure 1 and Table 1.

From the astrometric observations, it appears that the Orcus satellite is close 
to being pole-on and is in an essentially circular orbit.

We determine the orbit using a 
Powell $\chi^2$ minimization scheme to find the optimal orbital 
parameters. We first attempt to fit a purely circular orbit in which
the 5 free parameters are semimajor axis, 
orbital period, inclination, longitude of the ascending node, and mean anomaly.
The best fit has
a $\chi^2$ value of 9.5, or a reduced $\chi^2$ for 13 degrees of freedom 
(9 sets of $x$, $y$ coordinates
minus 5 orbital parameters) of 0.73, indicating an excellent fit to the
model (and perhaps slightly over estimated error bars). 
Expanding the model to allow an eccentric fit gives a best
fit eccentricity of 0.002 and 
a slightly lower reduced $\chi^2$ of 0.52. Given the excellent fit to the
circular orbit and the only slight improvement when eccentricity is
allowed, we conclude that 
no evidence for a non-circular orbit is found.

To determine uncertainties in the individual parameters,
we perform 1000 iterations of circular orbit fit optimization where we
add gaussian noise with $\sigma$ equal to the measurement uncertainties
of the position measurements and solve for new orbital parameters.
We define the $1 \sigma$
uncertainties on the
parameters to be the range containing the central 68\% of the data. 
To estimate an upper limit to the eccentricity, we perform
an additional 1000 iterations allowing a fully eccentric fit and
take the $1 \sigma$ upper limit to eccentricity to be the point higher
than 68\% of the data. We find an upper limit of 0.0036 to the
eccentricity of the orbit.
Table 2 gives the ecliptic orbital elements of the satellite orbit.

The circular orbit gives a system mass of $6.32\pm0.05\times 10^{20}$ kg, or about 3.8\% the mass of the largest known dwarf planet, Eris.

As with most KBO satellite orbits, astrometric
 data have not been obtained over a long
enough time interval to break the geometric plane-of-sky degeneracy, thus
two similar solutions can be found. The second solution has a reduced $\chi^2$ of 0.75 compared to 0.73 for the nominal solution, indicating that both are
excellent fits. The two solutions give nearly identical orbital distances and periods and thus identical masses. Both solutions and their uncertainties are
shown in Table 2. For both solutions, the orbit of Vanth is highly inclined
compared to the orbit of Orcus, with an angle between the pole vectors of
73 or 109 degrees, and nearly pole-on to the sun, with an angle between the pole vector and the sun-Orcus line of 29 degrees or 31 degrees.

\section{Visible photometry of Orcus and its satellite}
The unusual spectral properties
 of the sallelites of Haumea \citep{2006ApJ...640L..87B,  2009ApJ...695L...1F}
have been
used to argue that these satellites cannot have been captured from the typical
Kuiper belt population. Indeed, the deep water ice absorption features
seen on the satellites of Haumea have been argued to be a consequence of
formation by a giant impact on a previously differentiated body. Similar deep
water ice absorption features on Charon \citep{2000Sci...287..107B} suggest that such 
spectral features could be a general signature of formation in a giant impact.

An examination of the spectral properties of the satellite of Orcus could
likewise help to reveal its origin. The satellite is, however, significantly
closer to the primary than is the outer satellite of Haumea, thus ground-based 
spectroscopy to determine the surface composition
of the satellite is virtually impossible. 
We resort, instead, to HST spectrophotometry as a proxy for spectroscopy.
\citet{2009ApJ...695L...1F}
demonstrated that such spectrophotometry could detect the presence of
deep water ice absorption on the satellites of Haumea. HST can likewise easily 
resolve the Orcus system and provide separate photometry of the two components.

Observations of the optical and infrared colors of Orcus and its satellite were 
obtained with WFPC2 and NICMOS on HST on 5 Dec 2007 and 11 Nov 2008, 
respectively. 
Photometry was performed using the same PSF-fitting as described above for 
astrometric fitting of the points, and the results are shown in Table 3. 
Uncertainties are obtained from the scatter of the individual measurements.

Colors of Orcus and its satellite are shown in Figure 2 for both the
native instrument magnitude system and an approximate conversion to Johnson $V$ - Cousins $I$. For the conversion, we construct a series of linearly reddened
solar spectra and use the IRAF package {\it synphot} to calculate both sets 
of colors. We also use the best fit reddened spectrum to approximately
convert the F606W magnitudes to Johnson $V$ and F814W magnitudes to Cousins
$I$ for both objects. While the color conversion should be quite accurate, 
we assume that the accuracy of the single band conversion is no
better than 5\%. With the derived $V$ band magnitudes of Orcus and Vanth of
$19.36\pm0.05$ and $21.97\pm0.05$ and assuming the phase function 
for Orcus measured by \citet{2007AJ....134..787S} we obtain absolute magnitudes of 
Orcus and Vanth of $H_V= 2.27\pm0.05$ and $4.88\pm 0.05$, respectively. 
The combined absolute magnitude of $H_V=2.17\pm 0.05$ agrees well with 
the measured value in $R$ band of $1.81\pm0.005$ \citep{2007AJ....134..787S}and the 
solar color of $V-R=0.37$.

For a 
comparison, similarly obtained colors of Haumea and its satellites are also
shown in Figure 2. 
While Haumea and its satellites shows the characteristic neutral color and
clear signature of deep water
ice absorption at 1.6 $\mu$m, the satellite of Orcus has a moderately red
spectral slope at optical wavelenths and appears flat across the infrared, 
spectral characteristics that are typical in the Kuiper belt \citep{2008AJ....135...55B,2009Icar..201..272G}.

Unlike the satellites of Haumea, the visible-infrared spectrum of the satellite
of Orcus does not argue 
against a capture origin. Unfortunately, without a clearer
understanding of the origin of the deep water ice absorption features on
the satellites of Haumea and on Charon, we cannot conclusively argue that the
lack of such spectral features on the satellite of Orcus is a compelling argument
that the satellite must have been captured. 

Regardless of the origin of the satellite, 
the difference between the visible colors of Orcus and its satellite is 
a stark contrast between the very strong correlation between primary and
satellite colors observed by \citet{2009Icar..200..292B}. 
Indeed, no other measured satellite-primary 
pair appears so discrepent. Benecchi et al. argue that the satellite-primary 
color correlation suggests that KBO colors are primordial and a function of 
formation location. If satellite capture only
occurs within a limited zone then satellites and primaries should have 
similar colors. The discrepent colors of Orcus and Vanth could be used to
argue, therefore, that Vanth is not a captured satellite. A compelling
argument could be made, however, that larger KBOs (at least those not large
enough to maintain methane) are preferentially bluer, like Orcus, thus the
color of Orcus could reflect evolution rather than origin, while the color
of Vanth would be closer to the colors of the objects of the region in which
Orcus and Vanth formed. While no statistically significant evidence for such
a size-color correlation has been reported \citep{2008ssbn.book...91D}, 
this lack of
significance reflects as much the very small numbers of large objects as
it does the observations. We are thus reluctant to make a conclusion based
on the different optical colors of Orcus and Vanth.

The different colors certainly indicate different surface types and
suggest the possibility that Orcus and Vanth could have quite different
albedos.

\section{The size and density of Orcus}

To further characterize the properties of the Orcus system, we obtained far-IR
observations of the thermal emission from Orcus using the Multiband Imaging 
Photometer (MIPS) instrument on the Spitzer Space Telescope. 
These, and similar
observations of other TNOs, were described in \citet{2008ssbn.book..161S}.
Here we briefly re-summarize the data analysis and photometric methods.

Data were taken in the MIPS 24$\mu$m and 70$\mu$m channels using the photometry 
observing template. Observations were acquired on 30 and 31 May 2007.
Data were processed through the MIPS instrument team analysis pipeline,
resulting in absolutely-calibrated images. Between the 2 observations, Orcus
moved 35\farcs. We took advantage of that motion to subtract off background
sources in the images before performing photometry. This sky subtraction
provided a factor of $\simeq 2$ improvement in signal-to-noise ratio in both bands.
Finally, we coadded the sky-subtracted images from the two epochs in the co-moving
frame, producing a single final image in each MIPS band.

Photometry was measured using circular apertures with radii 5\farcs\ and 14.8\farcs\ 
at 24 and 70$\mu$m respectively, and appropriate aperture corrections applied
to the result \citep{2007PASP..119.1019G,2007PASP..119..994E}. 

We calculate flux values of $0.378\pm 0.03$ mJy at 24 $\mu$m and $25.0\pm1.4$ mJy
at 70 $\mu$m, where the fluxes are from the average of 4 measurements and
the uncertainties are determined from the scatter in the observations.
Including the 4\% and 8\% calbration uncertainties at 24 and 70 $\mu$m, our final
photometric measurements are $0.378\pm 0.03$ and $25.0\pm2.4$ mJy at
24 and 70 $\mu$m.

We approach modeling of thermal emission using the same suite of thermal
models \citep{LebSpen} with a different philosophy than
has been done previously \citep{2008ssbn.book..161S, 2009Icar..201..284B}. 
Results are similar, but our analysis has the potential to 
give more physical insight into the parameters used.

In the standard thermal model (STM) the surface temperature at 
each location on the body is assumed to be in instantaneous 
thermal equilibrium between incoming sunlight and outgoing thermal emission.
The total thermal emission of the body can be found by integrating the emission
over the entire surface. The other main end-member used in modeling is the
isothermal latitude model (ILM) in which each surface element is in radiative
equilibrium with the average sunlight that is seen over the course of a 
rotation, leading to a surface temperature which is a function of latitude.

Even for the lowest plausible values of thermal inertia,
objects at the $\sim$40-50K temperatures in the Kuiper belt radiate
so slowly that their surfaces do not have time to radiate
significant heat over typical Kuiper belt rotation periods. A simple order-of-magnitude
demonstration of this fact can be found by 
calculating the total temperature
change over a rotation in a volume one thermal skin depth deep. 
For any plausible thermal parameters the temperature change is only
a fraction of a degree. We thus conclude that the ILM is the only
physically plausible models. The ILM and STM are
identical when the rotation pole points directly at the sun.

The other main parameters used in these themal models are $q$, the phase 
integral, which relates the optical albedo to the Bond albedo, and $\eta$, 
the beaming factor, which is a simple correction to the total amount of
energy radiated in the sunward direction, usually assumed to be caused
by surface roughness, but which can be taken as a generic correction factor
to the assumed temperature distribution. For asteroids of known sizes, 
\citet{1986Icar...68..239L} found $\eta$ to be approximately 0.75, a correction
which agrees well with measurements of icy objects in the outer solar
system objects
\citep{1982Natur.300..423B,1982Icar...52..188B}. As no beaming
factor has been measured for any object in the Kuiper belt we allow this
factor to be a free parameter in our models, though we keep in mind that a 
value of approximately 0.75 appears to be widely applicable in the solar
system.

The Bond albedos or phase integrals of Kuiper belt objects have also never
been measured, but we find that the phase integrals of the large icy 
satellites of Uranus, whose infared spectra are quite similar to the
infrared spectrum of Orcus, all have measured phase integrals between
0.45 and 0.65 \citep{1990Icar...84..203B}. We use this as a representative
range for the phase integral of Orcus with the caveat that if the albedo
of Orcus turns out to be significantly different from those of the 
Uranian satellites we should revisit the appropriateness of the surface
analogy.

Finally, in our models we fix the pole position of Orcus
to be aligned with the derived pole positions of of the orbit of Vanth. 
If Vanth were created in a collision and tidally evolved outward
the poles would be expected to be aligned. For other formation
scenarios alignment of the rotation and satellite pole positions
is still a likely outcome, but not required (see below).

With these choices of parameters, we calculate a suite of thermal
models encompassing a full range of albedos, phase integrals, and beaming
parameters. Each model is scaled to a size to give it the total $V$ band
absolute magnitude of 2.17$\pm$0.06 from the WFPC2 data, then
fluxes in the MIPS 24 and 70 $\mu$m
bands are calculated. Rather than attempting thermal color corrections, as has
been done by all previous KBO thermal modelers, we instead directly
 integrate the
flux in the full bands using the published 24 and 70 $\mu$m filter functions
so that we can directly compare our measured fluxes to the models.

To assess the quality of the fit of any model, we define 
$$
\sigma=[ (F24_{\rm model}-F24_{\rm measured})^2/\sigma_{24}^2+(F70_{\rm model}-F70_{\rm measured})^2/\sigma_{70}^2], $$
where $F24_{\rm model}$ and $F24_{\rm measured}$ are the 24 $\mu$m modeled and 
measured fluxes, $\sigma_{24}$ is the uncertainty in the 24 $\mu$m measured
flux, and $F70_{\rm measured}$, $F70_{\rm modeled}$ and $\sigma_{70}$ are 
identical parameters at 70 $\mu$m.

Figure 3 shows the 24 and 70 \micron fluxes as well as the suite of
thermal models that fall within 1$\sigma$ of the measurements. Figure 4
shows a contour plot of $\sigma$ values as a function of diameter, albedo,
and beaming parameter. The best fit and full 1$\sigma$ range of uncertainty
to the diameter and geometric albedo are $940\pm70$ km and $0.28\pm0.04$, respectively. The
1$\sigma$ range of acceptable beaming parameters varies from 0.70 to 0.88.
This range encompasses the expected value of approximately 0.75 
previously found to be appropriate for asteroids and icy satellites,
giving confidence to the results. The derived albedo is also similar to
that of the icy satellites of Uranus that were used for phase integral
analogs, further bolstering confidence. 

The total thermal flux also includes the flux from Vanth. In the F606W filter,
Vanth is $2.54\pm0.01$
 magnitudes fainter than Orcus, suggesting a diameter 3.2 times
smaller for similar albedos. Given the moderately high albedo of Orcus,
however, and the dissimilar colors of Orcus and Vanth, we entertain the
possibility that Vanth could have an albedo as low as a factor of two
lower than Orcus. For similar albedos the combined sizes which yield
an equivalent surface area as that of a 940 km single object are 900 km and
280 km for Orcus and Vanth, respectively. Assuming similar densities (which,
again, might not be the case for this pair), the mass ratio is 33. If
Vanth has an albedo half of that of Orcus, 
the diameters would instead be 820 km and 640 km
with an equal-density mass ratio of only 2.1! (Owing to the extrmely
wide range of possible size and mass ratios, we have not explicitly 
recalculated thermal models for pairs of dissimilar objects, so true
best fit model sizes will differ slightly.)

The single-body size and mass of Orcus combine to give a density of $1.45 \pm 0.3$ g cm$^{-3}$, though partitioning the thermal emission to two bodies raises
the best fit density to 1.6 g cm$^{-3}$ for the identical albedo assumption
and to 1.5 g cm$^{-3}$ for the low albedo for Vanth. Regardless of the
partitioning, Orcus and Vanth appear to have a density in the intermediate
range between the high and low density clumps of Kuiper belt objects.

\section{The Origin of the Orcus Satellite System}
\subsection{Collision plus tidal evolution}
The close circular orbit of Vanth demands circularization
through tidal evolution. Such tidal evolution
would be expected from a collisionally formed
satellite that evolved outward.  

In this scenario, Vanth reaccumulated from
post-giant impact debris in an orbit just beyond the
Roche limit, and
the
post-collision Orcus was rotating rapidly
 from the collision. 
Tidal torques then transfered
angular momentum from Orcus' spin to the satellite orbit
and expanded its orbit outward. 
Eventually, the orbit of Vanth will expand
until the spin rate of both Orcus and Vanth
reaches the orbital frequency and Orcus and Vanth attain the double
synchronous state, as has occured for Pluto and Charon.

If Orcus and Vanth are currently in the double synchronous state,
we can put limits on their mass ratio. 
Assuming that both objects rotate with the 9.53 day
period of the orbit, the 
total angular momentum of the current system matches
a system where the primary is 
rotating at breakup and the secondary orbits at the
Roche radius if the mass ratio is approximately 8, well within the range
of allowed mass ratios. If the post-impact Orcus was not rapidly 
rotating and had a spin period more like the
$\sim$10 hour periods of typicals KBOs, however, the orbit would be
doubly synchronous with a mass ratio of 32.

Assuming the initial fast spin of Ocrus, for mass ratios 
larger than 8, the system is still evolving outward towards the double
synchronous state, while for mass ratios lower than 8 the current orbital configuration
contains too much angular momentum to have been achieved through outward tidal evolution
of an initial compact orbit.

We can estimate the timescale to tidally evolve to the current orbital distance as
$$
t={2\over 39} {({a\over r})}^5 {m_p\over m_s} {1\over n} {Q\over k_2},
$$
where $a$ is the satellite semimajor axis, $r$ is the radius of the primary,
$m_p/m_s$ is the primary to secondary mass ratio, $n$ is the orbital frequency, and $Q$ and
$k_2$ are the tidal dissipation factor and Love number, respectively \citep{MandD}. 
For a mass ratio of 8, this timescale becomes 6000 yr $\times (Q/k_2)$ or about
24000 yr $\times (Q/k_2)$ for a mass ratio of 32.
Using order-of-magnitude estimates of $Q=100$ and $k_2=0.005$ from \citet{MandD}, we
estimate a synchronization time scale of 
100 - 400 Myr for the system. Even within the large
uncertainties, this estimate is comfortably smaller than the age of the solar system, 
consistent with the idea that the system evolved to a synchronous state and
then stopped.
The eccentricity damping timescale
is even faster, so that a circular orbit is expected due to tidal dissipation. 

Is there evidence for a double synchronous state? 
Unfortunately, if the spin axis of Orcus is aligned 
with the orbit-normal (as expected for a tidally evolved system), 
the pole-on nature of the system implies that observing rotational 
photometric variability from Orcus is nearly impossible. Alternatively, a
robust detection of a spin rate faster than the 9.5 day period of the
Vanth orbit would conclusively demonstrate that this state has not been
achieved.

\subsection{Inward tidal evolution}
While outward tidal evolution can reproduce the current state of the
Orcus-Vanth system, tidal evolution can also proceed inwards from a 
more extended orbit if Vanth were 
captured rather than formed in a collision. For such an evolution
to produce the currently observed system, however, the capture would
have had to start with a highly unusual orbit. Because tidal evolution
conserves the total angular momentum of the system, and because
the total angular momentum of a system with a more extended orbit is
dominated by the orbital angular
momentum, this orbital angular momentum must stay approximately constant or increase.
Thus, the quantity $a(1-e^2)=a(1-e)(1+e)$ must be conserved, so the initial pericenter of the orbit $q_{\rm init} = a_{\rm init}(1-e_{\rm init})$ must be {\it smaller} than the current pericenter
as long as $a$ has decreased. While no general theoretical understanding
of the initial $a$ and $e$ distributions of captured Kuiper belt satellites
exists, invoking an initial orbit with an initial pericenter even closer than
the 19 primary radii of the current orbit appears implausible. We thus reject
this scenario as possible formation mechanism for the Orcus-Vanth system.

\subsection{Kozai cycles with tidal friction}
While an initially low pericenter orbit 
appears implausible, secular perturbations due to the sun can sometimes
create large eccentricies and low pericenters through Kozai oscillation
\citep{2009ApJ...699L..17P}
Kozai oscillations or Kozai cycles are large variations in eccentricity
and inclination which can occur when the relative inclination between the heliocentric orbit
and the binary mutual orbit exceeds a critical value (typically
$40^\circ < i < 140^\circ$) 
These oscillations
cause the eccentricity and inclination of the KBO binary to exchange on long timescales 
while keeping $\sqrt{1-e^2} \cos i$ constant.
The Orcus-Vanth system has a mutual inclination of either 73 or 109$^\circ$
putting it within the range where Kozai oscillations could have
effected it.

In the Kozai cycles scenario, Vanth is captured by Orcus on a typical
high semimajor axis moderate eccentricity orbit which happens to have
a high mutual inclination. This high mutual inclination allows
Kozai cycling to trade eccentricity and inclination, leading to
phases at which the orbit is extremely eccentric but the semimajor axis
remains unchanged. The pericenter of the satellite orbit can thus drop
down into the range where, like discussed above, significant tidal
evolution can finally occur at which point the semimajor axis
of the orbit can shrink dramatically.

In Ragozzine (2009)\footnote{Note to editor/reviewer: this reference to Darin Ragozzine's thesis (which
is available online through ADS will change to Ragozzine and Brown, submitted/in press when the paper is submitted/in press}
 we present detailed models of this
evolutionary path showing that Kozai cycles with tidal friction 
can produce orbital parameters similar to the Orcus-Vanth
system from a variety of starting conditions.

\section{Conclusions}
Two dynamical scenarios appear equally plausible for explaining the
formation of the Orcus-Vanth binary. The orbital characteristics of
Vanth are consitent with formation in a sub-catastrophic giant 
impact such as that that is thought to have formed the Pluto-Charon
system\citet{2005Sci...307..546C}. Indeed, with Orcus in a nearly identical
heliocentric orbit as Pluto suggesting an analogy between Pluto-Charon formation
and Orcus-Vanth formation is pleasing.

The Orcus-Vanth system could be equally well explained, however, by
capture, Kozai cycling, and subsequent tidal evolution (which {\it cannot} 
explain Pluto-Charon due to the presence of additional 
circular coplanar satellites
in that system). While the 2.54 magnitude
brightness difference between Orcus and Vanth is moderately extreme compared
to other binaries which are thought to have formed by capture \citep{2008ssbn.book..345N}, if Vanth has an albedo lower than Orcus by a factor of
two, as seems plausible, the equivalent magnitude difference is only 1.8, 
closer to values observed in typical systems.

We have hypothesized that perhaps all collisionally formed satellites
should share similar spectral characteristics, with the neutral
colors and deep water ice absorptions found in the Haumea system and on
Charon. While the presence of these unusual spectral features on Vanth
would have been significant evidence for a collisional origin it is
difficult to argue that their absence rules out collisional origin.
Without a significantly more detailed understanding of the physics of
icy body collsions and subsequent surface evolution we are unwilling
to assert that collision could not have formed a satellite with
a surface like Vanth.

Finally, we have noted that the largest KBOs with small presumably
collisionally formed satellites all have densities higher than those
of the smaller satellites. The uncertainty in the
density measured for Orcus, unfortunately, perfectly straddles the range
between those low density captured objects and higher density objects like
Pluto and Charon (though it is inconsistent with the extremely high
densities of objects like Haumea and Quaoar). 

The origin of the Orcus-Vanth binary remains uncertain. Future observations
which may help to constain this origin include more precise measurements
of the size of Orcus, to determine whether Orcus fits into the high
or into the low densities found in the Kuiper belt (or elsewhere), and 
a deep search for additional coplanar
satellites, analogous to Nix and Hydra, which
would firmly rule out a capture origin.

 {\it Acknowledgments:}
This research has been supported by grants from STScI and SSC and
through the NASA Earth and Space Science Fellowship program.

\begin{figure}
\plotone{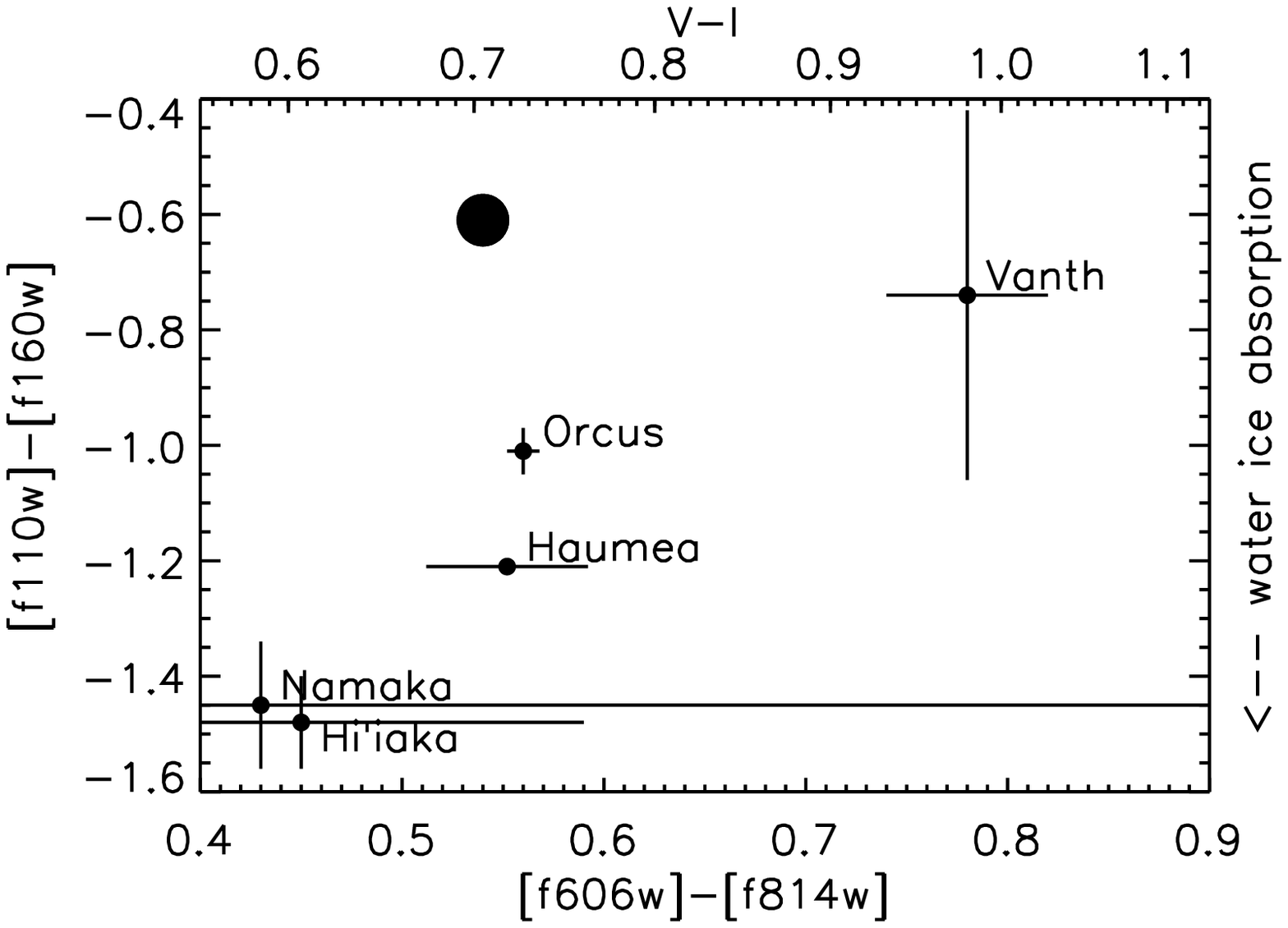}
\caption{Visible and infrared photometry of the Orcus and Haumea systems.
While the visible and infrared colors of Haumea and its satellites
Hi'iaka and Namaka are all consistent with having deep water ice
spectral features, the colors of Orcus and of Vanth are not. 
Orcus is known to have moderate water ice absorption, which 
can be seen in the moderate blue [f110W]-[f160W] colors. The large
circle shows solar colors in these filter systems.}
\end{figure} 

\begin{figure}
\plotone{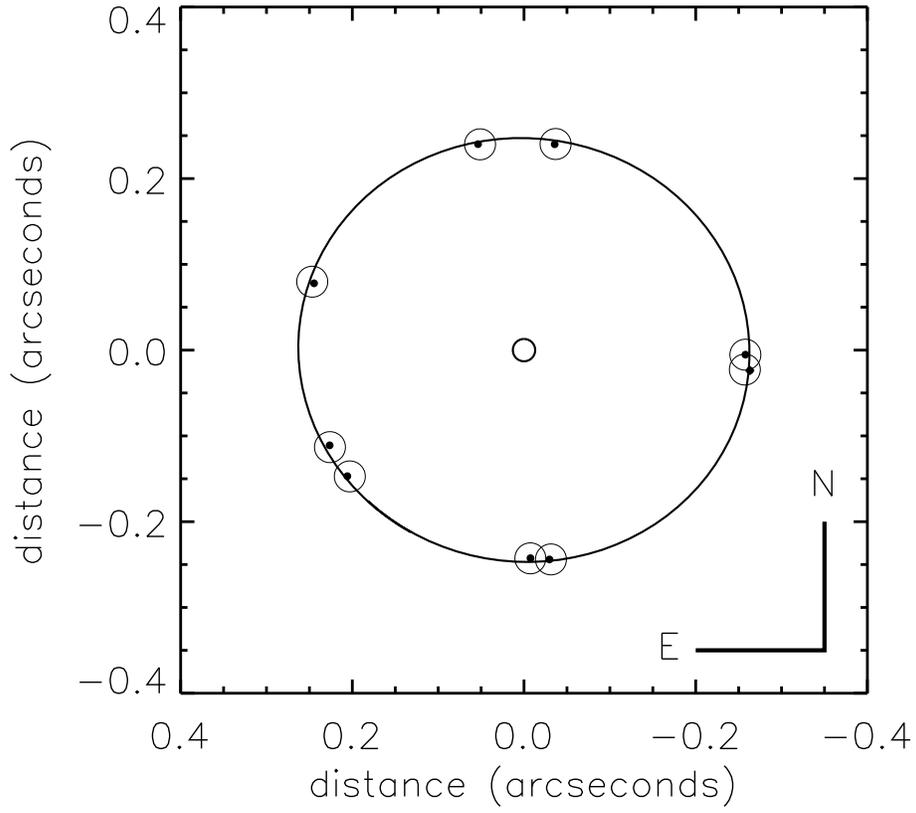}
\caption{The orbit of Vanth. The data points are shown with their error
bars, while predicted positions from the best fit models are shown as
large circles. Orcus is shown approximately to scale at the center.}
\end{figure}

\begin{figure}
\plotone{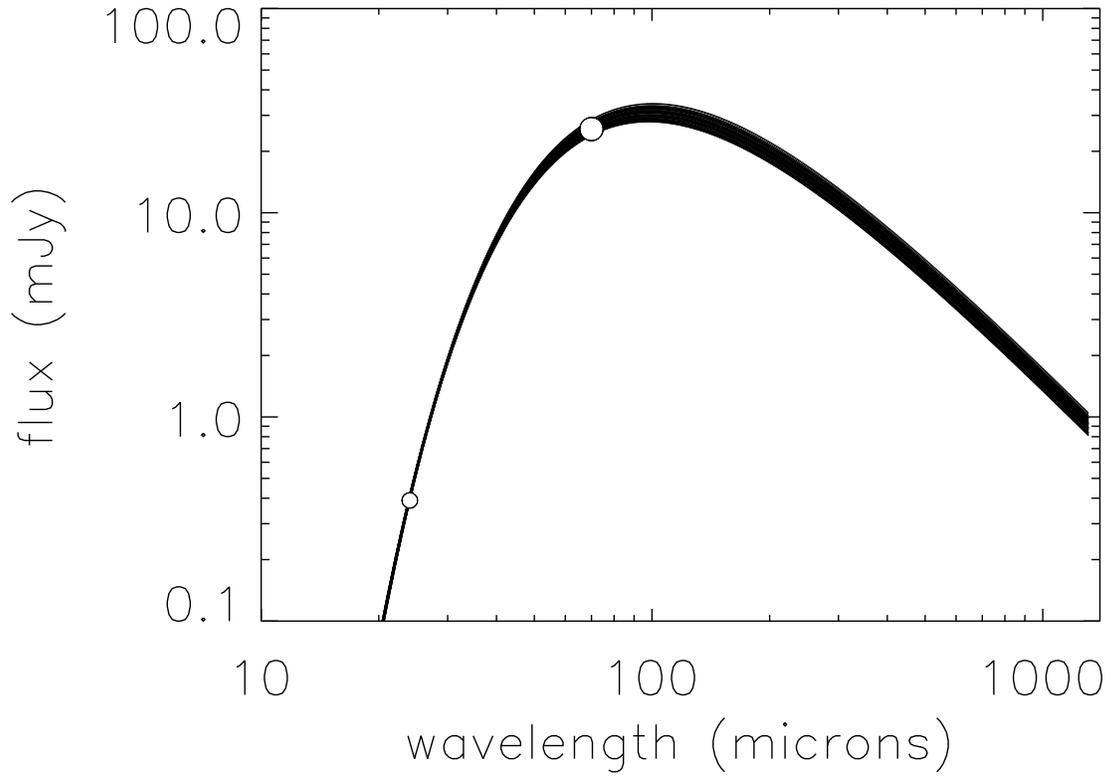}
\caption{Thermal models for Orcus. The 24 and 70 $\mu$m points from Spitzer
are shown with 1$\sigma$ error bars. The suite of models which provide
a 1$\sigma$ or better fit to these two data points is shown.}
\end{figure}

\begin{figure}
\plotone{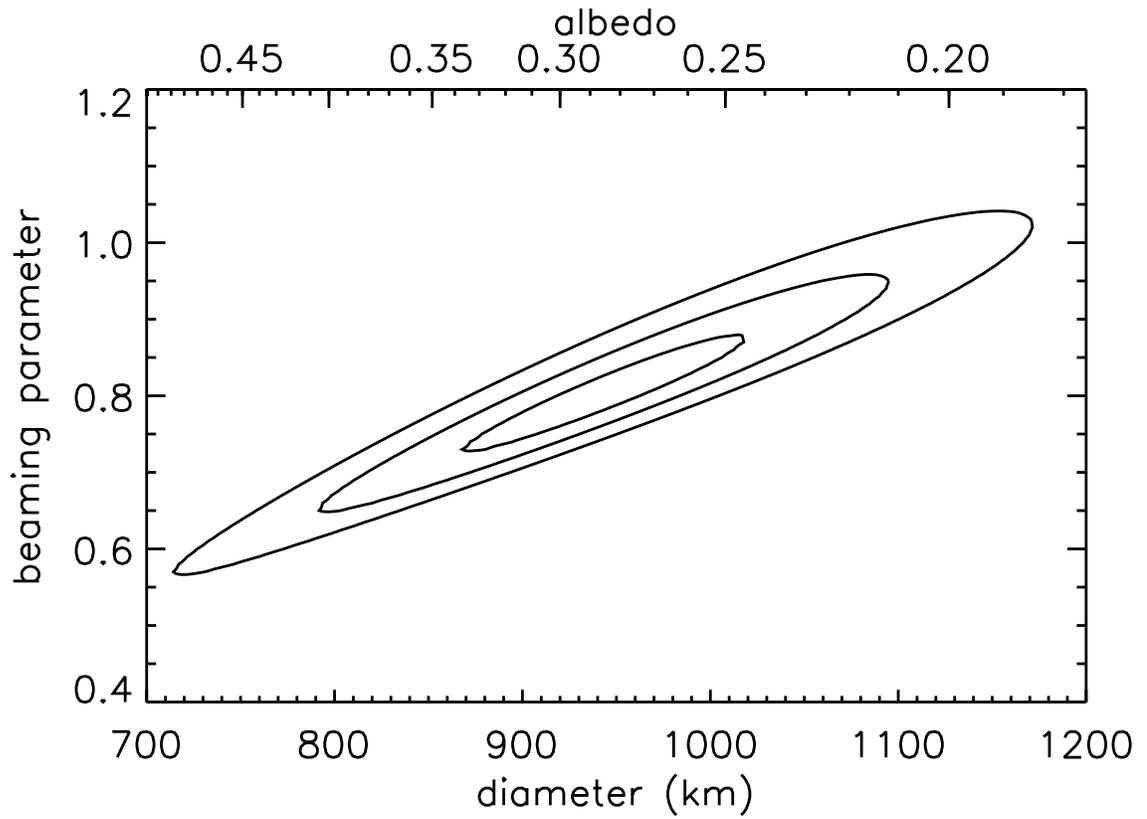}
\caption{Error contours for thermal model fits. Thermal fluxes from a large
grid of possible models with variable albedo, phase integral, and albedo
were calculated. The best fits have 
beaming parameters within the range of expected values.}
\end{figure}

\begin{table}
\begin{center}
\caption{Separation of Orcus and its satellite}
\begin{tabular}{lrrl}
\tableline\tableline
date &  RA offset & dec offset & instrument \\
(UT) & (mas) & (mas) \\
2453687.164 &$    206\pm 3$ &$    -147\pm 3 $& ACS \\
2454039.869 &$    226\pm 4$ &$   -111\pm 3 $& ACS \\
2454043.866 &$   -258\pm 1$ &$-005\pm 1 $& ACS \\
2454051.079 &$ -006\pm 1  $ &$ -243\pm 1 $& ACS \\
2454055.589 &$  -036\pm 1 $ &$   240\pm 1  $& ACS \\
2454065.646 &$   053\pm 2 $ &$   240\pm 1 $& ACS \\
2454079.836 &$  -030\pm 1 $ &$  -244\pm 1 $& ACS \\
2454415.793 &$   -263\pm 2$ &$  -024\pm 5 $& NICMOS \\
2454439.280 &$    245\pm 3$ &$   078\pm 4 $& WFPC2 \\

\tableline
\end{tabular}
\end{center}
\end{table}

\begin{table}
\begin{center}
\caption{Orbital parameters\tablenotemark{a}}
\begin{tabular}{lll}

\tableline
& solution 1 & solution 2\\
reduced $\chi^2$ & 0.73 & 0.75 \\
semimajor axis & $8980\pm 23$ & $8985\pm 24$ km \\
period & $9.5393\pm0.0001$ & $9.5392\pm 0.0001$ days \\
inclination & $90.2\pm 0.6$ & $305.8\pm 0.6$ degrees \\
longitude of ascending node & $50.0\pm 0.6$ & 249.4$\pm $0.4 degrees\\
mean anomaly & $143.1\pm 0.02$ & $316.6\pm 0.2$ degrees\\
epoch (defined) & JD 2454439.280 \\
\tableline
\end{tabular}
\tablenotetext{a}{Relative to J2000 ecliptic}
\end{center}
\end{table}

\begin{table}
\begin{center}
\caption{Photometry of Orcus and Vanth}
\begin{tabular}{rllll}
\tableline\tableline
date & instrument & filter/band & Orcus & Vanth \\
2454439.280&	 HST/WFPC2 & F606W	& $19.186\pm 0.006$ & $21.73\pm 0.01$ \\
	&	& F814W & $18.62 \pm 0.02$ & $20.90 \pm 0.02$ \\
	&	& Johnson $V$& $19.36\pm0.05$ & $21.97\pm0.05$ \\
	&	& Cousins $I$& $18.63\pm0.05$ & $20.94 \pm 0.05$ \\
2454415.793&	HST/NICMOS & F110W	& $20.64\pm 0.04$ & $23.3\pm 0.3$\\
	&	& F160W & $21.65\pm 0.03$ & $24.0 \pm 0.02$\\
	&Spitzer/MIPS & 24 & 0.378$\pm$0.3 mJy & (unresolved) \\
	& 	     & 70 & 25.0$\pm$2.4 mJy & (unresolved) \\
\tableline
\end{tabular}
\end{center}
\end{table}

\end{document}